

 \magnification\magstep 1
 \vsize=8.5truein
 \hsize=6truein
 \voffset=1 truecm
 \hoffset=1truecm
 \baselineskip=20pt
 \nopagenumbers

\def\hef{$^4\!$He }
\def\het{$^3\!$He }
\def\rr{{\bf r} }
\def\rp{{\bf r'} }

{ }

\vskip 2 truecm

\centerline{\bf ATOMIC AND MOLECULAR IMPURITIES IN \hef
CLUSTERS}

\vskip 2 truecm

\centerline{F. Dalfovo}

\centerline{\it Dipartimento di Fisica,  Universit\`a di
Trento, 38050 Povo, Italy}

\vskip 3 truecm

\noindent {\bf Abstract.}\ \ {\it A density functional
theory is used to predict the binding energy of atomic and
molecular impurities (Ne, Ar, Kr, Xe, Li, Na, K, Rb, Cs,
and SF$_6$) in the center of  \hef clusters, in the limit of
zero temperature and for zero angular momentum states.
The size dependence of the binding energy, from
small clusters to the bulk liquid limit, is investigated.
The behaviour of the \hef density near the impurity
is also studied.}

\vskip 2 truecm

PACS: 36.40; 67.40

\vfill\eject

\centerline{\bf I. INTRODUCTION}

\bigskip

Helium clusters are produced in supersonic nozzle beam
expansions (see Ref.[1] for a recent review).
They are expected to behave like  liquid droplets down to
zero temperature. At low temperature they are strongly
affected by quantum correlations and, eventually, become
superfluid. Several attempts have been made  in the
last years in order to   get experimental evidence of
superfluidity in helium clusters. Due to the weakness of
the helium-helium interaction a direct characterization of
pure helium clusters is very difficult.  A more promising
approach is the investigation of atomic
and molecular impurities attached to the helium clusters
[2,3]. The interpretation of the experimental data is still
limited by the lack of quantitative theories. The quantum
mechanical description of the static and dynamics of helium
clusters has been tackled by several authors [4-10], but
very little is known about the behaviour of impurities
[9-11].

In this work we calculate the energy and the density
distribution associated with impurity states in the center
of helium clusters at zero temperature and zero angular
momentum. We employ a density functional method, which was
developed in the last decade [5,12-14] in
the context of inhomogeneous states of liquid helium. With a
relatively small numerical effort it provides quantitative
predictions which are close to the results of {\it ab
initio} Monte Carlo calculations in the case of small
clusters, and can be extended to large clusters, up to the
bulk liquid limit. Quantum correlations between helium
atoms are accounted for by means of a phenomenological
density dependent interaction. The impurity is included as
an external potential in which the helium density adjusts
to minimize the energy. Rare gas atoms, alkali atoms, and
SF$_6$ molecule are considered. Accurate impurity-helium
potentials are taken from the literature [15-18].

In Section II we introduce the density functional
formalism, emphasizing the main physical features and
discussing the approximations made in the treatment of the
impurity.  In Section III we present the results of the
calculations. Section IV is a short summary  of the main
results and a discussion about open problems and future
work.

\bigskip
\bigskip

\centerline{ \bf II. \  METHOD}

\bigskip

In a density functional theory one writes the total energy
of the many-body system as a functional of the one-body
density $\rho (\rr)$, in the form
$$
E = \int \! d\rr \ {\cal H} [\rho] \ \ \ . \eqno (1)
$$
Under certain conditions the minimization of $E$
with respect to $\rho$  is equivalent to the solution of
the many-body Schr\"odinger equation [19].  In
general, however, the exact form of the functional, whose
minimum is located at the true equilibrium density of the
system, is not known {\it a priori}. Thus suitable
phenomenological functionals are introduced; they yield
approximate results, whose quality depends on how the
relevant symmetries and correlations are included in the
starting functional form.

A first systematic description of helium clusters in the
framework of density functional is the one of Ref.~[5].
The theory was built in such a way that the
compressibility and surface tension of liquid helium were
reproduced. The results for the cluster energy and density
profile  were close to the ones of quantum variational
calculations [4].  The functional of Ref.~[5] has
been recently extended to include the effect of the finite
range helium-helium interaction [13]. The new
functional has the form:
$$
{\cal H}_\circ =  {\hbar^2 \over 2m}  (\nabla
\sqrt{\rho(r)}\ )^2 + {1\over 2} \int \!  d\rp \
\rho(r)  \rho(r') V(|\rr-\rp|) +  {c
\over 2} \rho(\rr)  (\bar \rho_{\rr})^{1+\gamma} \ \ \
. \eqno (2)  $$
The first term in the sum is a quantum
pressure;  it  corresponds to  the zero temperature kinetic
energy of a noninteracting Bose system. The second term
contains a two-body interaction $V$, which is the
Lennard-Jones interatomic potential, with the standard
parameters  $\alpha =2.556$ \AA \ and $\varepsilon=10.22$
K,  screened at a distance $h\!=\!2.377$~\AA \  with a power law,
as follows
$$ V (x) =
\cases{
4 \varepsilon \left[ \left(  {\alpha \over x} \right)^{12}
- \left( {\alpha \over x} \right)^6  \right] , &if $x\ge
h$;\cr V(h) \left( {x \over h} \right)^4 , &if $x < h$.\cr
} \eqno (3)
$$
The last term in Eq.~(2) accounts for
short range correlations  between atoms. In particular it
contains the effect of the hard core  part of the
interatomic potential. Its form follows the idea of the
"weighted density approximation", used mainly in the study
of classical  fluids. The  weighted density $\bar \rho$ is
the average of $\rho (\rr)$ over a sphere with radius $h$.
The parameters $c=1.04554 \times 10^7$ K
\AA$^{3(1+\gamma)}$, and $\gamma=2.8$, together  with the
screening length $h$, are the only three parameters of the
theory; they are fixed to reproduce  the equation of state
of the bulk liquid. As discussed in ref.[13] the functional
${\cal H}_\circ$ corresponds to a  mean-field description, which
incorporates  phenomenologically the  effects of a
finite-range interaction, with the correct long range
behaviour, as well as  of short range correlations. Like
the functional of Ref.~[5] it reproduces the equation of
state, the compressibility and the surface tension of bulk
liquid \hef. Moreover it reproduces the behaviour
of the static response function, which is peaked at the
roton wavelength. This ensures the inclusion of
localization effects, which are crucial to predict the
freezing transition at high pressure [20], the layer
structure in helium films [21,22], and the shell structure
near impurities [14].

The effect of an impurity can be included in the density
functional by adding a potential term
$$
{\cal H} = {\cal H}_\circ +  V_I (\rr) \rho
(\rr) \ \ \ , \eqno (4)
$$
where $V_I(\rr)$ is the helium-impurity potential. This
corresponds to treat the impurity as a classical object
in a quantum liquid. Since the  zero point motion of
the impurity is not accounted for,  the validity of Eq.~(4)
is restricted to impurities heavier than the helium atomic
mass $m$. So, we do not discuss \het or hydrogen
impurities. For them a full quantum mechanical treatment is
necessary [10,11].  In this work we consider rare gas
atoms, alkali atoms and SF$_6$ molecules. For most of them
the effect of zero point motion is surely negligible. Only
for Li and Ne it gives rise to sizeable effects,
but, as we will discuss later,  the potential contribution
is expected to be still dominant.

Combining Eqs.~(1), (2)  and (4) the minimization with respect to
$\rho$ yields the following Euler-Lagrange equation:
$$
\left[ -{\hbar^2 \over 2m} \Delta  + U(\rr) +
V_I(\rr) \right] \sqrt{\rho} = \mu \sqrt{\rho}
\ \ \ , \eqno (5)
$$
where $\mu$ is the helium chemical potential, while
$U(\rr)$ is  the Hartree self consistent field derived from
functional (2) [13,14].
Equation~(5) can be solved in bulk helium as well as in  clusters
of given particle number $N$. The solution yields both the
density profile and the energy of the system. The impurity
chemical potential can be extracted as  difference
between the energy of the cluster with and without the
impurity (X)
$$
\mu_I= E[ {\rm X (He)}_N] -E[{\rm (He)}_N] \ \ \ . \eqno
(6)
$$
In bulk this is equivalent to
$$
\mu_I = \int\! d\rr \ \left({\cal H} [\rho(\rr)] - \mu
\rho(\rr) \right) \ \ \ . \eqno (7)
$$

The solution of Eq.~(5) in bulk helium as been already
discussed by Pavloff [14], who considered Na, Cs, Xe and
Ba$^+$ as impurities and used the  same density functional for
liquid helium.  The results for Xe were in good agreement
with previous variational calculations by K\"urten and
Ristig [23]. For  Cs the agreement was only
qualitative, mainly because  the role of the elementary
diagrams neglected in the calculations of Ref.~[23] is
expected to be more important for Cs than for Xe. Another
test on the density functional method is the evaluation of
the energy of one atom of \hef, considered as a classical
impurity in the rest of the liquid. The calculation, with
helium-helium interaction in place of $V_I$, yields an
energy of about $-23$ K. This is in good agreement with
independent estimates of the potential energy per particle
in bulk helium. The difference between this value and the
total energy per particle, $-7.15$ K, is an estimate of
the zero point kinetic energy. The kinetic energy for
different impurities scales approximately as the mass
ratio and the square of the radius of the bubble created
by the impurity inside the liquid. For the lightest
impurities considered in this work (Li and Ne)  the
resulting  zero point energy is no more than  $3 \div 4$ K,
which is much less than the potential energy.

In the present work we calculate the impurity
chemical potential and the helium density near the
impurity, in  bulk liquid and clusters, using accurate
impurity-helium interaction [15-18]. We  consider only
impurity states in spherical symmetry. Thus Eq.~(5) becomes
a  one-dimensional equation in the radial coordinate,
which can be solved numerically with a standard iterative
procedure (a few minutes on a RISC-CPU for each run).
In the case of helium clusters, the assumption of spherical
symmetry  implies impurity states in the center of the
cluster. The inclusion of possible bound states outside the
cluster (surface states) will be a subsequent step, which
requires further numerical efforts in solving Eq.~(5) with
anisotropic density.

\vfill\eject

\centerline{\bf III. \ RESULTS }

\bigskip

\centerline{\bf III.A \  Impurity in Bulk Liquid}

\bigskip

We take the mixed rare gas van der Waals potential
by  Tang and Toennies [16], the alkali-helium potential
by  Patil [18], and the spherical approximation of the
SF$_6$-helium potential by Pack et al. [15], with a
modified value of the potential depth as in Ref.~[17].
We solve Eq.~(5) in bulk liquid helium by imposing the
asymptotic value of $\rho$ equal to the uniform liquid density
at fixed external pressure. The results for the impurity
states are summarized in Table~I. In the first column
the impurity chemical potential (at zero temperature
and zero pressure) is given. One notes that it is negative
for rare gas atoms and for SF$_6$, while it is positive for
alkali atoms. This reflects the different structure of
$V_I$; the alkali-helium potential has a more extended
repulsive core and a weaker attractive tail than the
rare-gas and the SF$_6$ potential. As a consequence the
rare gas impurities, as well as SF$_6$, tend to compress
the helium atoms in shells around the impurity, in a
region of large and negative $V_I$, with a gain in
energy. On the contrary, the alkali atom pushes the helium
atoms far away without changing significantly the helium
density; this corresponds to a  cost in energy.
Density profiles  are shown in Fig.~1. The height of the
first peak in  helium density is also given in the second
column of Table I (the equilibrium density of the uniform
liquid is $0.0218$ \AA$^{-3}$).  The energy systematics and the
structure of the density profiles are very similar to the situation
of helium films on solid substrates [21,22], where one finds
a transition from wetting to non wetting as a function of
the substrate-helium potential parameters.

An interesting quantity is the number of helium atoms in the
first shell, close to the impurity (third column in Table~I).
The most attractive case is the one of SF$_6$, where we find
approximately $25$ atoms in the first shell, corresponding
to a solid  {\it snowball} surrounding the impurity.  This
{\it snowball} makes the impurity  state less sensitive to the
external pressure, since the behaviour of state is dominated by the
local shell structure. The chemical potential of  alkali atoms
is much more pressure dependent. It tends to increase with
the external pressure, because it costs more energy to
create a {\it bubble} in a liquid under pressure.  For instance, the
chemical potential of Na increases from $49$ K to $74$ K in passing
from zero to $5$ bars, while the one of Kr decreases from $-243$ K to
$-250$ K  in the same range.
It is worth noticing that the existence of a shell structure near
an impurity, as in the case of SF$_6$ and rare gas atoms, has
important consequences in the dynamic properties of the system.
In particular it is expected to change considerably the effective
mass  of the impurity as well as the structure of the velocity field
of  helium in the  surrounding region.

The accuracy of our predictions depends
on two main aspects: the quality of the impurity-helium
interaction used as input and the quality of the density
functional ${\cal H}_\circ$. Indeed one can take from
the literature different impurity-helium potential for
the same kind of impurity. The resulting chemical potential
scales approximately as the well depth of $V_I$, and
the height of the first peak in $\rho$ slightly changes.
But, at the present level of understanding, the precision in the
impurity-helium potential is not crucial.  For this  reason we have
also chosen the isotropic approximation for the SF$_6$-helium
potential instead of the true anisotropic one. As concerns the
quality of the density functional, one can test it
by comparison with quantum Monte Carlo calculations, whenever
available, as done in different contexts [5,11,13,22,24].

\bigskip
\bigskip

\centerline{\bf III.B \ Impurities in the center of helium
clusters}

\bigskip

The solution of Eq.~(5) without the impurity potential
corresponds to the case of pure \hef clusters. The
results  for the energy systematics and for the density
profiles are very close to the  ones of Ref.~[5], where a
simpler density functional was used. This  confirms the
fact that for smoothly varying density the relevant
quantities in the theory are the compressibility and the
surface tension, which are both well reproduced by the two
functionals.

We solve Eq.~(5) with and without impurities. The
difference in energy provides the impurity chemical
potential. A major advantage of the density functional
method is that it works easily even with quite large
clusters, up to $N=5000$ particles or more. So   one can
test the asymptotic convergence to the bulk liquid limit,
predicting  the size dependence of the results on a wide
range of $N$.

In Figs.~2 and 3 we show the impurity chemical potential
as a function of  $N$. Again there is a significant
difference between alkali atoms and the other impurities.
The alkali atoms are more sensitive to the size of
the cluster and reach the asymptotic value of the chemical
potential in bulk  more slowly. This effect can be
understood by looking at Fig.~4, where the density
profile for a cluster with Na and Kr is
compared with the pure helium cluster. One notes that the
rare gas atom does not modify the external
structure of the cluster. As a consequence its chemical
potential is fixed mainly by the local distortion of the
density near the center, and  the energy is almost the
same as in bulk liquid. On the contrary the Na atom
pushes the helium atoms outside, working against the
local pressure of the liquid and increasing the surface
area. The net effect is an increase of the impurity chemical
potential with respect to the bulk value. This is also
in agreement with the pressure dependence of
the chemical potential in bulk, which is stronger for
alkali atoms.

In all cases the chemical potential bends towards zero in
the limit of small clusters. This happens when the
first shells of atoms near the impurity  become
partially occupied. In the case of SF$_6$ and rare gas
the attractive impurity-helium potential dominates on the
helium-helium correlations; thus a lack of atoms in the
first shells, in the potential well of the impurity,
implies an increase of  energy. Viceversa in the case of
alkali atoms the helium-helium correlations are more
important and a decrease in  helium density  implies a
decrease in energy.

Typical density profiles are shown in Fig.~5. One notes that
the first shell of atoms near the rare gas
impurity is deformed only for very small $N$.
This is a clear sign that the impurity prefers to tie
the helium atoms around itself, i.e., to stay in
the center of the cluster with a relatively large
binding energy.  This effect is even stronger
in the case of  SF$_6$.  On the contrary, the  alkali
atoms seem not to bind to the center of the cluster
for any value of $N$. The energy of the impurity-cluster
system is higher when the impurity is in the center than
when it is far outside. This does not exclude {\it a
priori} the possibility to capture alkali atoms on helium
droplets. In fact, since at large distance the relative
interaction is attractive, a local  energy minimum may
still exist on  the surface of the cluster.

Monte Carlo calculations for small clusters with
impurities are also becoming available. In Fig.~6 we
compare the density profile for a cluster of $111$
particles with a SF$_6$ molecule in the center. The dashed
line is the result of Diffusion Monte Carlo calculations
[10], and the solid line is the prediction of our density
functional theory, using the same impurity-helium
potential. The agreement is very good. The small
difference in the first peak height is well within the
expected accuracy of the present theory.

\bigskip
\bigskip

\centerline{\bf IV. \ CONCLUSIONS}

\bigskip

We have done density functional calculations for rare gas
atoms, alkali atoms and SF$_6$ molecule in the center of
helium clusters. We have discussed the size dependence of
the impurity chemical potential and of the helium density
profile  on a wide range of particle number, up to the bulk
liquid limit. Our  results strongly support the
existence of bound states in the center of a cluster for
rare gas atoms and SF$_6$, but not for alkali atoms. For
SF$_6$ in small clusters  the predictions of the density
functional theory are in good agreement with the ones
of recent Diffusion Monte Carlo calculations [10].

In view of the current debate about  the location of
impurities on clusters [2,3] further theoretical work is
needed. A first possibility is to drop the spherical
symmetry in Eq.~(5), to allow for impurity states on the
surface of the clusters, as well as to study the cluster-impurity
interaction as a function of the relative distance.
This makes the numerical computation  heavier, without any
substantial change in the theory.
A second point is the inclusion of non zero angular momentum,
which is expected to favour surface impurity
states [2]. The inclusion of  velocity field and vorticity
in the framework of the density functional theory has been
already discussed for bulk liquid helium [24]. The
treatment of the velocity field in the cluster geometry is
expected to be more difficult. The problem deserves
certainly further investigations, being  directly related to the
concept of superfluidity in finite systems.

\vskip 1.5 truecm

{\bf ACKNOWLEDGEMENTS}

This work was partially supported by INFN, gruppo collegato
di Trento.

\vfill\eject

{\bf REFERENCES.}

\bigskip

\item{1.} Toennies, J.P.: In:  {\it Proceedings
of the International School of Physics E. Fermi on "Chemical
Physics of Atomic and Molecular Clusters", Course CVII}, p.597.
Scoles, G. (ed.). Amsterdam: North Holland 1990

\item{2.} Schutt, D.L.: Thesis, Princeton 1992;
Goyal, S., Schutt, D.L., Scoles, G.:
Phys. Rev. Lett. {\bf 69} 933 (1992);
Goyal, S., Schutt, D.L., Scoles, G.: Chem. Phys.
Lett. {\bf 196}, 123 (1992);
Goyal, S., Schutt, D.L., Scoles, G.: J. Phys. Chem. {\bf 97},
2236 (1993)

\item{3.} Scheidemann, A., Toennies, J.P., Northby, J.A.:
Phys. Rev. Lett. {\bf 64}, 1899 (1990);  Scheidemann, A.,
Schilling, B., Toennies, J.P.: Preprint, G\"ottingen 1993

\item{4.} Pandharipande, V.R., Zabolitzky, J.G., Pieper, S.C.,
Wiringa, R.B., Helbrecht, U.: Phys. Rev. Lett. {\bf 50},
1676 (1983); Pandharipande, V.R., Pieper, S.C., Wiringa, R.B.:
Phys. Rev. B {\bf 34}, 4571 (1983);
Pieper, S.C., Wiringa, R.B., Pandharipande, V.R.: Phys. Rev. B
{\bf 32}, 3341 (1985)

\item{5.} Stringari, S., Treiner, J.:  J. Chem. Phys. {\bf
87}, 5021 (1987)

\item{6.} Casas, M., Stringari, S.: J. Low Temp. Phys. {\bf 79},
135 (1990)

\item{6.} Sindzingre, P., Klein, M.L., Ceperley, D.M.:
Phys. Rev. Lett. {\bf 63}, 1601 (1989)

\item{7.} Rama Krishna, M.V., Whaley, K.B.: Phys. Rev.
Lett. {\bf 64}, 1126 (1990); Rama Krishna, M.V., Whaley, K.B.:
J. Chem. Phys. {\bf 93}, 746 (1990); Rama Krishna, M.V., Whaley, K.B.:
J. Chem. Phys. {\bf 93}, 6738 (1990)

\item{8.} Chin, S.A., Krotscheck, E.: Phys. Rev. Lett.
{\bf 65}, 2658 (1990); Chin, S.A., Krotscheck, E.:
Phys. Rev. B {\bf 45}, 852 (1992)

\item{9.} Barnett, R.B., Whaley, K.B.: J. Chem. Phys. {\bf
96}, 2953 (1992)

\item{10.} Whaley, K.B.: Preprint, Berkeley 1993;
Barnett, R.B., Whaley, K.B.:  (in preparation)

\item{11.} Dalfovo, F.: Z. Phys. {\bf D 14}, 263 (1989)

\item{12.} Stringari, S.,and Treiner, J.:   Phys. Rev. {\bf B
36}, 8369 (1987)

\item{13.} Dupont-Roc, J., Himbert, M.,  Pavloff, N.,
Treiner, J.: J. Low  Temp. Phys. {\bf 81}, 31 (1990)

\item{14.} Pavloff, N.:  Thesis, Orsay 1990

\item{15.} Pack, R.T., Piper, E., Pfeffer, G.A., Toennies, J.P.:
J. Chem. Phys. {\bf 80}, 4940 (1984)

\item{16.} Tang, K.T.,  Toennies, J.P.: Z. Phys. D {\bf
1}, 91 (1986)

\item{17.} Scoles, G.:  Int. J. of Quantum Chemistry: Quantum
Chemistry Symposium {\bf 24}, 475 (1990)

\item{18.} Patil, S.H.: J. Chem. Phys. {\bf 94}, 8089 (1991)

\item{19.} Hohenberg, P., Kohn, W.:  Phys. Rev. {\bf 126},
B864 (1964); Kohn, W., Sham, J.:  {\it ibid} {\bf 140},
A1133 (1965); Mermin, N.D.:  {\it ibid} {\bf 137}, A1441
(1965)

\item{20.} Dalfovo, F., Dupont-Roc, J., Pavloff, N.,
Stringari, S., Treiner, J.: Europhys. Lett. {\bf 16}, 205
(1991)

\item{21.} Pavloff, N., Treiner, J.: J. Low Temp. Phys.
{\bf 83}, 331 (1991)

 \item{22.} Cheng, E., Cole, M.W.,  Saam, W.F., Treiner, J.:
Phys. Rev. Lett. {\bf 67}, 1007 (1991); Cheng, E., Cole, M.W.,
Saam, W.F., Treiner, J.:Phys. Rev. B {\bf 46}, 13967 (1992)

\item{23.} K\"urten, K.E., Ristig, M.L.: Phys. Rev. B
{\bf 27}, 5479 (1983);  K\"urten, K.E., Ristig, M.L.:
Phys. Rev. B {\bf 31}, 1346 (1985)

\item{24.} Dalfovo, F.: Phys. Rev. B {\bf 46}, 5482 (1992)

\vfill\eject

\centerline{\bf FIGURE CAPTIONS}

\bigskip
\bigskip

\item{\bf Fig.~1\ } Helium density profiles near impurities in bulk
liquid, at zero temperature and zero pressure. The profiles for rare
gas impurities  correspond to Ne, Ar, Kr, and Xe in this order
starting from the profile closest to $r=0$. The profiles for alkali
impurities corresponds to Li, Na, K, Rb, and Cs in the same order.
The cordinate $r$ is the distance from the impurity.

\item{\bf Fig.~2\ } Chemical potential of rare gas atoms, and
SF$_6$ molecule, in helium clusters as a function of the helium
particle number.

\item{\bf Fig.~3\ } Chemical potential of alkali atoms in helium
clusters as a function of the helium particle number.

\item{\bf Fig.~4\ } Density profile for $300$ helium atoms around
Na and Kr impurity. The dashed line is the density profile of a pure
helium cluster with the same number of particles.

\item{\bf Fig.~5\ } Density profile for helium clusters with Na and
Kr impurities.  The number of helium particles is 20, 30, 50, 100,
300, 1000, and 3000 starting from the smallest one.

\item{\bf Fig.~6\ } Density profile for a cluster of 111 helium atoms
and one SF$_6$ molecule in the center. Dashed line: Diffusion Monte
Carlo calculations of Ref.~[10]; solid line: present calculation with the
same impurity-helium potential.

\vfill\eject

\centerline{\bf TABLE CAPTION}

\bigskip
\bigskip

\item{\bf Table I. } Results for the impurity state in bulk liquid
helium at zero pressure and zero temperature.  From left to right:
impurity type,  impurity chemical potential, maximum value of the
helium density, and  number of helium atoms in the first shell.

\vfill\eject

\settabs 5\columns
\+& {\bf TABLE I}&\cr

\vskip 4truecm

\+ &$\mu_I$ \   [K] & $\rho_{\rm max}$\ (\AA$^{-3}$) & $N_{\rm shell}$ &\cr

\smallskip

\hrule
\hrule

\smallskip

\+ Li      & $+$  40   & 0.0242 & --  &\cr
\+ Na      & $+$  49   & 0.0237 & --  &\cr
\+ K       & $+$  68   & 0.0227 & --  &\cr
\+ Rb      & $+$  69   & 0.0226 & --  &\cr
\+ Cs      & $+$  84   & 0.0220 & --  &\cr
\+ Ne      & $-$  39   & 0.0477 & 12  &\cr
\+ Ar      & $-$ 195   & 0.0593 & 17  &\cr
\+ Kr      & $-$ 243   & 0.0605 & 19  &\cr
\+ Xe      & $-$ 313   & 0.0595 & 21  &\cr
\+ SF$_6$  & $-$ 601   & 0.0820 & 25  &\cr

\smallskip
\hrule

\bye